# A black hole in a globular cluster

Thomas J. Maccarone[1], Arunav Kundu[2], Stephen E. Zepf[2], Katherine L. Rhode[3,4,5]

1. School of Physics and Astronomy, University of Southampton, Southampton, SO17 1BJ, UK
2. Department of Physics and Astronomy, Michigan State University, East Lansing MI, 48824
3. Department of Astronomy, Wesleyan University, Middletown CT, 06459
4. Department of Astronomy, Yale University, New Haven CT, 06520
5. NSF Astronomy and Astrophysics Postdoctoral Fellow

**Globular star clusters contain thousands to millions of old stars packed within a region only tens of light years across. Their high stellar densities make it very probable that their member stars will interact or collide. There has been considerable debate about whether black holes should exist in these star clusters[1,2,3]. Some theoretical work suggests that dynamical processes in the densest inner regions of globular clusters may lead to the formation of black holes of ~1,000 solar masses[3]. Other numerical simulations instead predict that stellar interactions will eject most or all black holes that form in globular clusters[1,2]. Here we report the X-ray signature of an accreting black hole in a spectroscopically-confirmed globular cluster in the Virgo Cluster giant elliptical galaxy NGC 4472. This object has an X-ray luminosity of about $4*10^{39}$ ergs/sec, making it brighter than any non-black hole object can be in an old stellar population. The X-ray luminosity varies by a factor of 7 in a few hours, ruling out the possibility that the object is several neutron stars superposed.**

We have compared the locations of point sources found by the X-ray Multiple Mirror-Newton (XMM-Newton) satellite with the positions of globular clusters that have been identified in optical imaging data[4] and confirmed with optical spectroscopy (Zepf et al. in prep). We have found a source at RA=12:29:39.7, Dec=+7:53:33.4 (J2000) which has an average X-ray luminosity of $2*10^{39}$ ergs/sec. Its optical counterpart (with a positional offset of about 0.4", determined by using the more accurate *Chandra* astrometry from ACIS-S observations on 12 June 2000) is a bright blue globular cluster, with V=20.99, B-V=0.68, V-R=0.38, (indicating a metallicity of 1/50 solar, using a color-metallicity relation[5]) and has a spectroscopically-determined radial velocity of 1477+-7 km/sec, clearly identifying it with NGC 4472. Its optical luminosity is about $7.5*10^5$ times that of the Sun, making it one of the most luminous globular clusters in NGC 4472. This X-ray source was also detected previously by ROSAT[6], but as the source lies 6.6 arcminutes (about 30 kpc) from the center of the galaxy, there have been no previous reports of searches for optical counterparts.

The globular cluster is a secure optical counterpart for the X-ray source, with a very low probability of being a chance superposition of a background active galactic nucleus (AGN) on a globular cluster. Our spectroscopically confirmed globular cluster sample includes 53 clusters in an annulus between 6' and 8' around the center of NGC 4472, yielding a density of $1.6*10^{-4}$ globular clusters per square arcsecond at this distance from the cluster center, yielding a low probability (0.005) of having a globular cluster within 1" of one of the thirty brightest X-ray sources in this galaxy. In fact, only one of the 30 brightest X-ray sources is in this annulus. The space density of AGN at least as bright as the source (with an absorbed 0.5-2.0 keV flux of $5.6*10^{-14}$ ergs/sec/cm$^2$) is about $5*10^{-7}$ per square arcsecond[7], yielding a probability of only $6*10^{-4}$ of having such a bright AGN within 1" of any of the spectroscopic globular clusters.

Furthermore, this object has a soft spectrum, and varies much more strongly at soft X-rays than at hard X-rays. Thus, if it were a background AGN, it would have to be a Narrow Line Seyfert 1 galaxy, and these objects comprise only about 15% of AGN[8]. Finally, the AGN would have to be faint enough optically that it would neither affect the globular cluster's colors nor introduce redshifted emission lines into its spectrum. This constraint is not easily quantifiable; however, the fact that the typical broadband optical to X-ray flux ratio for AGN is larger than the ratio of fluxes for the globular cluster and the X-ray source[10] makes it even less likely that the X-ray source is a background AGN.

The location in a globular cluster of a given X-ray source is not surprising - about half of all X-ray sources in elliptical galaxies are in globular clusters[10,11,12]. There are, however, two remarkable features to this object – its X-ray light curve shows very high amplitude variability and its X-ray spectrum peaks at lower energy X-rays than the spectra of typical X-ray sources in galaxies. During the first 10,000 seconds of the observation by the X-ray Multiple Mirror (XMM)-Newton satellite, the European Photon Imaging Camera PN light curve shows a count rate of about 0.04 cts/sec from 0.2 to 12 keV. Over the next ~10,000 seconds, the count rate drops by a factor of about 7 to about 0.006 cts/sec, where it remains for the next 60,000 seconds, when the observation ended (see figure 1). The same results are seen in the Metal Oxide Semiconductor cameras, but with lower count rates due to the detectors' lower effective areas. Because of this high amplitude variability, we have extracted the spectrum in two intervals, one corresponding to count rates above 0.04 cts/sec, and the other to count rates below 0.03 cts/sec. In the high count rate interval, the spectrum is well fit by a disk blackbody model with an inner disk temperature of 0.22 keV, and an inferred inner disk radius of about 4,400 km, and Galactic absorption of 1.67 x $10^{20}$ H atoms/cm$^2$. The spectrum of the low count rate interval is consistent with having the same underlying continuum model, but with an increased neutral hydrogen absorbing column density ($N_H$) of about $3*10^{21}$ H atoms/cm$^2$; the source varies by a factor of about 10 below 0.7 keV and is consistent with being constant above 0.7 keV. In each case, the underlying, unabsorbed luminosity is about $4.5*10^{39}$ ergs/sec – the Eddington limit for accretion of hydrogen onto a 35 solar mass object, or heavier elements onto a 15 solar mass object. Similar spectral variability (i.e. variations consistent with changing absorbing column density) has been reported in the past in the stellar mass black hole candidate V404 Cyg[13], although the changes in the absorption needed there were a few times larger than those reported here. The only other Galactic black hole X-ray binary which has spent a substantial amount of time above $10^{39}$ ergs/sec is GRS 1915+105, which has a foreground $N_H$ of about $5*10^{22}$ H atoms cm$^{-2}$ (ref 14), so we would not expect to be able to see $N_H$ variations in GRS 1915+105 which are as small as the ones we are reporting in the NGC 4472 source.

This large variability amplitude is obviously inconsistent with the idea that the X-ray emission in this source could come from a superposition of bright neutron stars – and in fact this variability is one of the only ways to prove an extragalactic globular cluster X-ray source is a black hole[15]. If the inner disk radius of the accretion disk around the X-ray source is 4,400 km, and this corresponds to the innermost stable circular orbit of a Schwarzschild black hole, then the black hole mass is at least about 400 solar masses, so it is plausible that this object is an accreting intermediate mass black hole. The key remaining issue is whether the best fitting inner disk radius really is the innermost stable circular orbit around the black hole.

Proposals exist for producing soft X-ray spectra at high luminosities from stellar-mass, rather than intermediate-mass, black holes. At very high accretion rates, slim discs (i.e. discs with scale heights of about 10% of their radii) should form[16] around black holes, and these should have very large, optically thick photospheres. The source spectrum we see is consistent with the emission expected from a 10 solar mass black hole accreting at a rate of about $3*10^{-7}$ solar masses per year in the context of such models[17,18]. The variability in the source discussed here is predominantly in the soft X-rays, in direct contrast to what was seen in the M101 source first suggested as an example of this mode of accretion[17] – in such outflow models, the variability should be due to changes in the size of the photosphere, not in the absorption column in front of it. On the other hand, since V404 Cyg did show a similar X-ray spectrum[19] and similar spectral variability[13], and at a luminosity only slightly lower, it seems quite plausible that the object in NGC 4472 is a super-Eddington stellar mass black hole; perhaps the M101 source is at a lower inclination angle than V404 Cyg and the NGC 4472 black hole, so it is less obscured by the outflow generated through the super-Eddington accretion. Thus the case for an intermediate mass black hole in this cluster is inconclusive, in contrast to the strong case for a black hole of some mass rather than a neutron star.

It has also been suggested that it may be possible to exceed the Eddington limit by a large factor in slim disc models[18]. The luminosity may be Eddington limited at every radius of an accretion disk. The actual brightness will then be the Eddington limit times the logarithm of the ratio of the mass accretion rate to the Eddington mass accretion rate under the assumption of standard radiatively efficient accretion. The unabsorbed luminosity is then 20-26 times the Eddington rate for hydrogen accretion onto a 1.4 solar mass neutron star. This yields an outer disk mass accretion rate of at least 5 solar masses per year. Additionally, slim discs cannot exist around neutron stars, at least in the present models, since it will not be possible to advect trapped photons in the case of a neutron star accretor. Thus, even in the high accretion rate slim disc case, the identification of this source as one with a black hole accretor is secure. The other existing mechanism for making neutron stars accrete far above their Eddington limit is polar accretion in X-ray pulsars, such as SMC X-1 which exceeds its Eddington limit by a factor of about 15(ref 20), but this works only in young stellar populations because magnetic fields in accreting objects decay over time. Therefore, it is highly unlikely an accretion-powered pulsar will be a globular cluster source.

This discovery allows us to place some limited constraints on models of formation and evolution of systems with accreting black holes in globular clusters. Since this object has been bright at least since the ROSAT observations in 1994[6], and was detected in a Chandra observation and an XMM observation separated by about 3 years, it seems clear that the outburst duration is at least a decade (or that the source is persistent). In the context of models with stellar mass black hole accretors in globular clusters, this object is most likely in a system with a red giant donor star in a long (i.e. at least one month) orbital period[21,22]. It has been suggested that there should typically be a single stellar mass black hole in most globular clusters, and that most of these black holes should be in wide binaries, leading to low X-ray bright duty cycles[15]. At present, with only a single confirmed black hole in NGC 4472's ~6,000 globular clusters[4] we can estimate that the duty cycle of bright X-ray emission must not be much less than 1/6,000, or 1/400 if one includes only the spectroscopically selected clusters. In either case, this is consistent with previous suggestions that the duty cycle as bright X-ray sources of these systems should be no more than about 1/1,000.
Attempts to understand the population synthesis and stellar evolution of accreting binaries with intermediate mass black holes have also been made[23-26]. The theoretical predictions both of the probabilities the black hole will capture a companion, and of the probability that a cluster will contain an intermediate mass black hole are highly uncertain. Therefore, it is difficult to place constraints on theory with only a single object which may or may not be of intermediate mass.

**Acknowledgements**

We wish to thank Elmar Koerding, Tom Dwelly, Sebastian Jester, John Salzer and Gilles Bergond for useful comunications.

**Author information**

Reprints and permissions information is available at npg.nature.com/reprintsandpermissions. The authors declare no competing financial interests. Correspondence and requests for materials should be addressed to TJM at tjm@phys.soton.ac.uk.

**Figure legend:**

Figure 1. The X-ray light curve of the globular cluster black hole candidate in NGC 4472, from the EPIC-PN detector. Count rate from 0.2-12 keV is plotted versus time since the start of the observation. The error bars are 1σ uncertainties. The data in the light curve come from an XMM-Newton observation of NGC 4472 which lasted about 100,000 seconds and took place on 1 January 2003. We have used the standard procedures for cleaning the XMM data to remove high background intervals (yielding about 80,000 seconds of good time) and to extract source spectra and light curves.

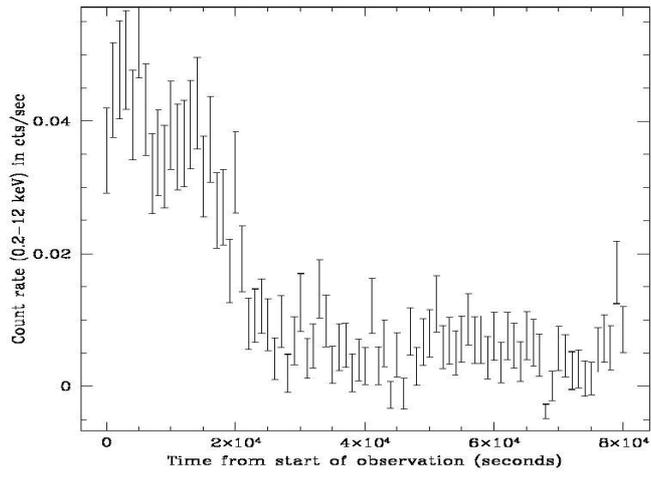

Title: The X-ray lightcurve of the NGC 4472 globular cluster black hole